\documentclass{moriond}

\def\be{\begin{equation}}
\def\ee{\end{equation}}
\def\bea{\begin{eqnarray}}
\def\eea{\end{eqnarray}}

\usepackage{xspace}
\usepackage{acronym}
\usepackage{subfigure}

\def\CW{CW\xspace}
\def\DM{DM\xspace}
\def\GW{GW\xspace}
\def\NS{NS\xspace}

\def\CWs{CWs\xspace}
\def\GWs{GWs\xspace}
\def\NSs{NSs\xspace}

\def\PBHs{PBHs\xspace}

\newcommand{\TFFT}{T_{\rm FFT}}

\newcommand{\Tobs}{T_{\rm obs}}

\newcommand{\Photo}{\includegraphics[height=35mm]{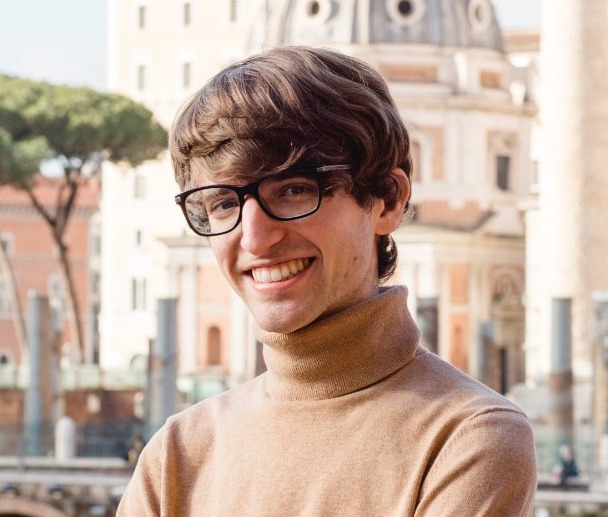}}

\begin{document}
\vspace*{4cm}
\title{Recent results from continuous gravitational wave searches using data from LIGO/Virgo/KAGRA's third observing run }

\author{ A.L. Miller, amiller@nikhef.nl \\ on behalf of the LIGO, Virgo and KAGRA Scientific Collaborations }
\address{Nikhef -- National Institute for Subatomic Physics,
Science Park 105, 1098 XG Amsterdam, NL \\ Institute for Gravitational and Subatomic Physics (GRASP),
Utrecht University, NL}

\maketitle\abstracts{
The third observing run of advanced LIGO, Virgo and KAGRA brought unprecedented sensitivity towards a variety of quasi-monochromatic, persistent gravitational-wave signals. Continuous waves allow us to probe not just the existence of canonical asymmetrically rotating neutron stars, but also different forms of dark matter, thus showing the wide-ranging astrophysical implications of using a relatively simple signal model. I will describe the major results from the numerous continuous-wave searches that were performed in O3, both inside and outside the LIGO/Virgo/KAGRA collaborations, and show how impactful to multi-messenger physics that they have been.}

\section{Introduction}

Continuous waves (\CWs) are quasi-monochromatic, persistent signals that could arise from deformed neutron stars (\NSs), inspiraling planetary-mass compact objects, annihilating boson clouds around rotating black holes, or from the direct interaction of dark matter (\DM) particles with standard model ones in gravitational-wave (\GW) interferometers\cite{Sieniawska:2019hmd,Piccinni:2022vsd}. Detecting any of these sources would provide deep insights into the \NS equation of state or the existence of dark matter, and open a new window into understanding the universe. Despite the different underlying physics in each of these sources of \CWs, all of the signals can be modelled as a sinusoid whose frequency slowly varies with time. For \NSs or boson clouds around black holes emitting \GWs coming from unknown sky locations, or for a \NS in a binary system, other modulations due to the Doppler motion of earth relative to the source, and orbital motion, must be accounted for.

Since \CWs last for durations much longer than the observation times of \GW detectors, \CWs are not one-off events: they would continue to be seen in \GW data, parameter estimates would continuously improve, and they could even be used to calibrate interferometer data. Furthermore, \CWs will be the operating model for a variety of signals in future ground- and space-based detectors, such as galactic white dwarf or intermediate-mass black hole binaries, or inspiraling sub-solar mass primordial black holes (\PBHs), respectively.

The long durations of \CW signals allows us to dig extremely deeply into the noise, obtaining strain sensitivities of $\mathcal{O}(10^3)$ times better than what can be achieved with compact binary mergers. But, when analyzing long stretches of data, different problems arise that are not necessarily present in searches for compact binaries coalescing: (1) frequent glitches, (2) noise non-stationarity, (3) gaps in data collection and (4) quasi-monochromatic instrumental noise lines, due to e.g. the resonance frequencies of the detectors' mirrors. Therefore, search algorithms must be robust against unpredicted noise disturbances, while remaining sensitive to \CWs. Additionally, most \CW searches -- with the exception of those that target known pulsars (see Sec. \ref{sec:targeted}) -- cannot rely on matched filtering, since the number of templates needed to cover the sky increases steeply with the observation time, i.e. $N_{\rm sky}\propto \Tobs^2 $, and the computing cost scales as $\Tobs^6$ \cite{Astone:2014esa}. In other words, for directed and all-sky searches for \NSs, the number of sky points to search over becomes of $\mathcal{O}(10^{13})$ when considering $\Tobs=1$ yr up to 2000 Hz; thus, a matched filtering analysis is intractable. Therefore, numerous pattern-recognition methods have been employed in the search for \CWs that divide the data into short time segments analyzed coherently, then combined incoherently. These so-called ``semi-coherent'' searches have numerous advantages compared to matched filtering: (1) immensely reduced computational cost, (2) robustness towards deviations in the signal model, and (3) less affected by the presence of noise disturbances. And, the sensitivity loss with respect to matched filtering for \CWs is only a factor of a few, due to the practical nature of performing \CW searches, e.g. the need to set thresholds, select a fixed number of candidates to follow up, etc \cite{Astone:2014esa}.

In the third observing run (O3) of advanced LIGO \cite{LIGOScientific:2014pky}, Virgo \cite{VIRGO:2014yos}, and KAGRA \cite{Aso:2013eba} (LVK), numerous analyses for \CWs were performed to search for isolated \NSs, dark matter and primordial black holes. In these proceedings, I describe the most recent results for searches for \CWs in O3, and comment on lessons that have been learned that will prepare us for O4.

\section{Neutron Stars}

\subsection{The Physics}


\GW emission could occur from \NSs in a variety of ways, if there is some deviation from spherical symmetry that causes a nonzero, time-dependent quadrupole moment. In the canonical case, an equatorial deformation, i.e. a ``mountain'' on the star's surface, could be sustained by the star's internal magnetic field, expected to be $1-10^4$ times stronger than the external one. This magnetic field could have become so potent after years of being buried through accretion, in the case of millisecond pulsars \cite{Mukherjee:2017rwl}. Furthermore, r-modes, a toroidal perturbation throughout the star's surface whose restoring force is the Coriolis force, could also be drive \GW emission \cite{owen1998gravitational}, especially in the case of PSR J0537$-$6910 \cite{Ho:2020vxt}. In all cases, the \GWs extract rotational energy from the star, spinning it down, or in the case of Scorpius X-1 or other X-ray binaries, balancing the spin-up induced by the accretion of matter from the companion onto the neutron star.

\GW emission could happen at $2f_{\rm rot}$, at $f_{\rm rot}$ and $2f_{\rm rot}$, or at $\sim 4/3f_{\rm rot}$, where $f_{\rm rot}$ is the star's rotational frequency, if the star exhibits free precession \cite{Jones:2001yg}. The first case is due to non-axisymmetric rotation of the neutron star, the second could be due to a pinned superfluid core interacting with the star's magnetic field, and the third could be due to r-mode emission \cite{Piccinni:2022vsd}.

It is easier focus on the $2f_{\rm rot}$ case; we can therefore write the \GW amplitude $h_0$ in terms of the size of the deformation:

\be
h_0 =1.0\times 10^{-25} \left(\frac{1{\rm kpc}}{d}\right) \left(\frac{f_0}{1000{\rm Hz} }\right)^2  \left(\frac{I_{\rm zz}}{10^{38}{\rm kg\cdot m^2} }\right)  \left(\frac{\epsilon}{10^{-7} }\right)
\ee
where $d$ is the distance from the solar system barycenter to the source, $I_{\rm zz}$ is the principle moment of inertia about the z-axis, and $\epsilon=\frac{I_{\rm xx}-I_{\rm yy}}{I_{\rm zz}}$ is the equatorial ellipticity, which can be interpreted as the ``height'' of a mountain if multiplied by the radius of a \NS.
Regardless of what physical mechanism is behind sustaining the \NS deformation, we can write an equation that describes all deterministic frequency shifts that a quasi-monochromatic \GW would experience:

\begin{equation}
f_{\rm obs}(t)=f_0+\dot{f}(t-t_0)+\frac{\ddot{f}}{2}(t-t_0)^2 + f_0\frac{\vec{v}(t)\cdot\hat{n}}{c}-f_0a_p\cos{\Omega(t-t_{\rm asc}})
\label{eqn:master}
\end{equation}
where $f_{\rm obs}$ is the observed \GW frequency at the detector at time $t$, $f_0$ is the source \GW frequency at a chosen reference time $t_0$, $\dot{f}$ is the spin-up or spin-down, $\ddot{f}$ is the second derivative of the \GW frequency, $\vec{v}(t)$ is the velocity of the detector with respect to the solar system barycenter, $\hat{n}$ is the sky position of the source, $a_p$ is the projected semi-major axis, $\Omega$ is the orbital angular frequency, and $t_{\rm asc}$ is the time of ascension.

The fourth term in this expression describes frequency modulations due to the relative motion of the earth and the source, which must be considered when we do not know, or are uncertain of, the position of the \NS. The last term here details the frequency change due to the orbital motion of a \NS around a companion object. 

%

In the following subsections, I will describe the, broadly speaking, three types of \CW searches performed for neutron stars: targeted, directed, and all-sky. Each one has its benefits and drawbacks, which I will explore in the context of the most recent LIGO/Virgo O3 results.

\subsection{Targeted searches}
\label{sec:targeted}

\emph{Targeted} searches look for pulsars whose existence has already been confirmed via electromagnetic observations. Currently, $\sim$ 3000 pulsars have been identified in the Galaxy, with rotational frequencies ranging from $\mathcal{O}(1-1000)$ Hz \cite{Manchester:2004bp}. Astronomers can measure very precisely the position, period, period derivative and binary orbital parameters, though the second period derivative is more uncertain, since pulsars continuously beam radio waves or X-rays at us.

From the timing solutions provided by astronomers, the so-called ``ephemerides'', we can derive the (almost) exact rotational frequency, spin-down, and second-order frequency derivative, and therefore have a complete model for how the \GW emission would appear in our data. We can then use matched filtering to correlate the ``template'' \GW emission with the data, and compute a signal-to-noise ratio that encodes how much our template matches the data.

In targeted searches, we typically compare the sensitivities of our searches to the amplitude of \GWs that would be emitted assuming that all of the star's rotational energy is converted into \GWs, i.e. the amplitude at the so-called "spin-down limit".

\be
h_{0,\rm sd} = 2.5\times 10^{-25}\left(\frac{1{\rm kpc}}{d}\right) \sqrt{\left(\frac{f_0}{1000{\rm Hz} }\right) \left(\frac{10^{-10}{\rm Hz/s}}{\dot{f}}\right) \left(\frac{I_{\rm zz} }{10^{38}{\rm kg\cdot m^2} }\right)}
\ee

\begin{figure*}[ht!]
     \begin{center}
        \subfigure[]{%
           \label{fig:msp_cons}
\includegraphics[width=0.48\textwidth]{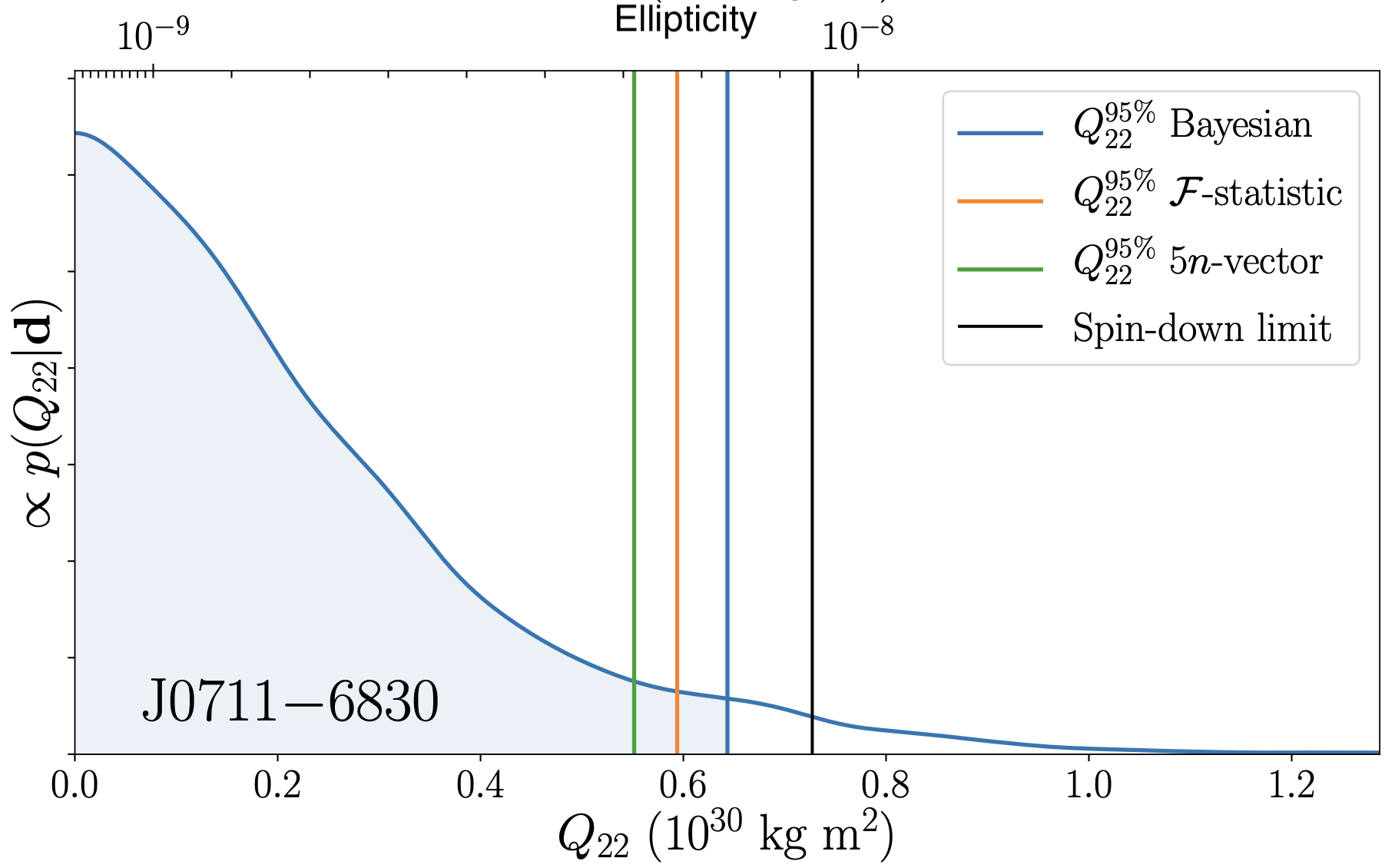}
        }
                \subfigure[ ]{%
            \label{fig:j0537}
            \includegraphics[width=0.42\textwidth]{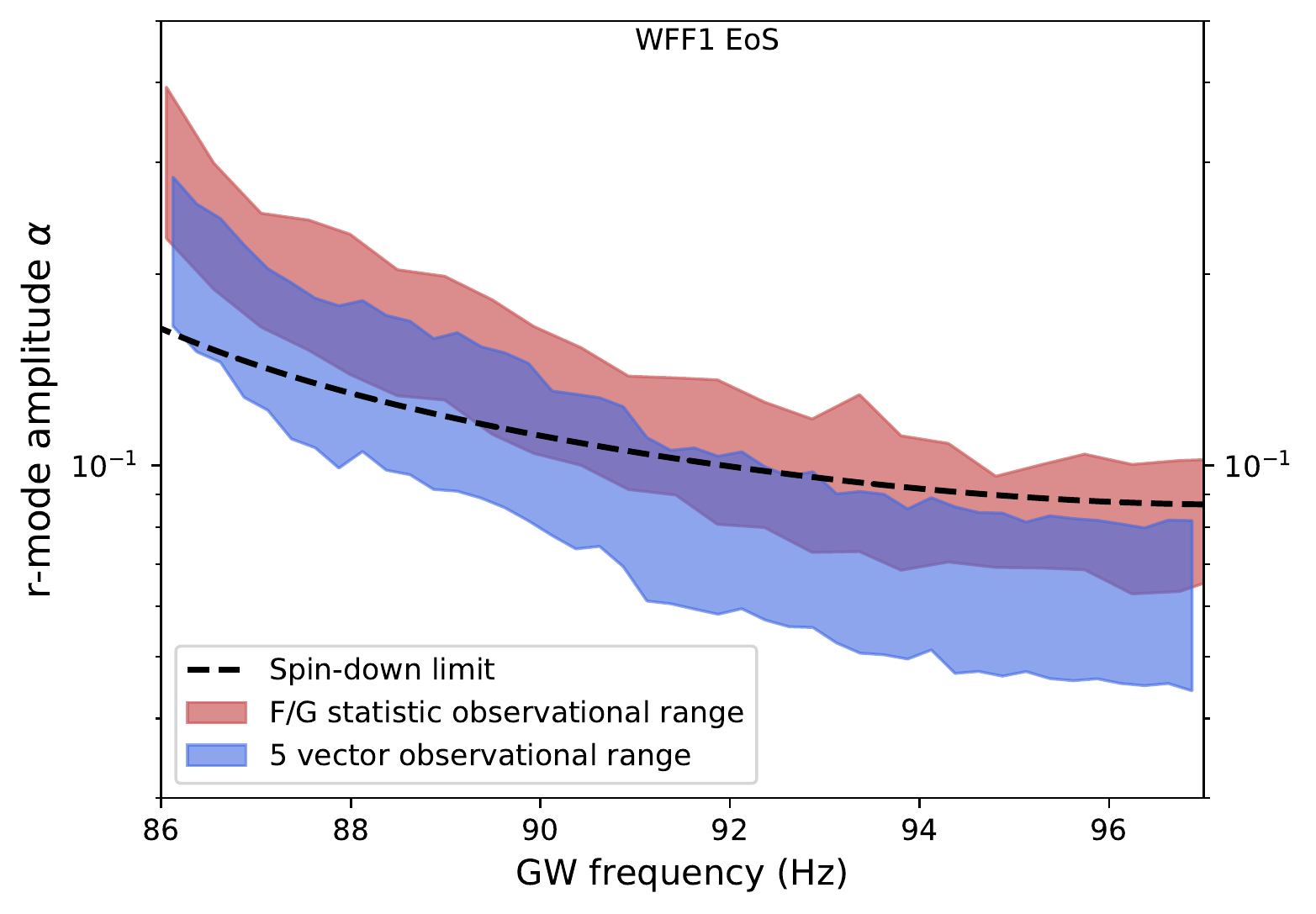}
        }\\ 
    \end{center}
    \caption[]{%
   (a) Posterior distribution on the ellipticity in the Bayesian search \cite{LIGOScientific:2020gml}. Each method's 95\% upper limit is drawn as a vertical line, and lies below the spin-down limit, indicating that the spin-down limit has been surpassed for this pulsar. (b) Upper limit on the r-mode saturation amplitude for two search methods for a particular equation of state, with the spin-down limit shown, for PSR J0537$-$6910 \cite{LIGOScientific:2021yby}. }%
     \label{fig:targeted}
\end{figure*}

In O3, multiple targeted searches were performed for both \GWs due to an equatorial deformation, and due to r-mode emission \cite{LIGOScientific:2021hvc,LIGOScientific:2020lkw,LIGOScientific:2021quq}. For the first time, the LVK collaborations surpassed the spin-down limit for millisecond pulsars (MSPs), which tend to be older, and are expected to have different evolutionary histories, than pulsars such as Crab and Vela \cite{LIGOScientific:2020gml}. The ellipticities constrained from MSPs are a few orders of magnitude smaller than those from Crab and Vela, implying that, at this point, a much smaller deviation from spherical symmetry would be required to sustain a deformation on MSPs compared to young pulsars. We show an example of upper limits computed in a Bayesian way \cite{LIGOScientific:2020gml} for MSP J0711$-$6830, in which each method produced an upper limit on ellipticity lower than the spin-down limit, in Fig. \ref{fig:msp_cons}.

The first search for r-modes from J0537$-$6910 was also performed in O3, using ephemerides provided by the NICER collaboration \cite{LIGOScientific:2021yby}. This search was motivated by the fact that the long-term behavior of the braking index $n$, a quantity related to the measured parameters $f_{\rm rot}$, $\dot{f}$, and $\ddot{f}$, tends towards $n=7$, indicating r-mode emission. Furthermore, this pulsar, colloquially known as the ``Big Glitcher'', glitches often (few times per year), is close ($\sim 50$ kpc away), is young $\mathcal{O}(1000)$ yrs old, and spins rapidly $f_{\rm rot}=62$ Hz. Fig. \ref{fig:j0537} shows constraints on the r-mode saturation amplitude $\alpha$ as a function of possible GW frequencies for a soft equation of state for two different methods used in this search, compared to the calculated spin-down limit, which has been surpassed for most frequencies shown.

Typically, the matched filtering searches do not explicitly deal with the glitches; they either (1) analyze periods before and after glitches, or (2) assume unknown phase offsets between electromagnetic and \GW phases and marginalize over them. However, searches for post-glitch \GWs can also be carried out, essentially meaning that the \CW signal turns off after a certain amount of time.
Furthermore, narrow-band searches were also performed at known targets, in which some mismatch between the predicted \GW frequency given the electromagnetically measured frequency parameters  is allowed. This kind of search is more computationally heavy than the targeted ones, since it must consider many more templates; however, it is robust against uncertainties in the measured electromagnetic parameters \cite{LIGOScientific:2021quq}. 

While targeted and narrowband searches are the most sensitive ones that we can perform, we expect that we are only observing a small fraction of the $\mathcal{O}(10^9)$ neutron stars that exist in our galaxy. Thus, we need to perform other kinds of analyses in order to truly probe the unknown, which will be detailed in the next subsections.

\subsection{Directed searches}

When we decide to look in a particular sky location, but we do not know any of the source parameters (or even how many sources could be there), we perform \emph{directed} searches. Examples include searches that target the Galactic Center \cite{KAGRA:2022osp}, supernova remnants, \cite{LIGOScientific:2021mwx}, and X-ray binaries \cite{LIGOScientific:2022enz}. Each of these places is expected to house neutron stars that we do not see electromagnetically, implying that a \GW detection would allow a new way to probe the \NS population.

In directed searches for isolated sources, the unknowns are $f_0$ and $\dot{f}$ ($\ddot{f}$ is not considered); however, the notion of the size of a particular location on the sky is important. Though $\dot{f}$ causes the smallest frequency variation in Eq. \ref{eqn:master}, the longer we observe for, the finer we can resolve the sky. Thus, we cannot analyze all data coherently, since even a ``fixed'' sky location (e.g. the galactic center) could require analyzing a huge number of sky points, at \emph{each} unknown $f_0$.

Thus, \emph{semi-}coherent methods had to be designed, ones that break the dataset of length $\Tobs$ into smaller chunks of length $\TFFT$, analyze these individual chunks coherently, then combine them incoherently (that is, with the loss of phase information).

The $\TFFT$ length is chosen as a compromise between computational cost, not losing too much sensitivity compared to the matched filter, and being robust against unmodeled physics that may cause stochastic or deterministic deviations to the model in Eq. \ref{eqn:master}.

\begin{figure*}[ht!]
     \begin{center}
        \subfigure[ ]{%
            \label{fig:gr15sco-x1}
            \includegraphics[width=0.6\textwidth]{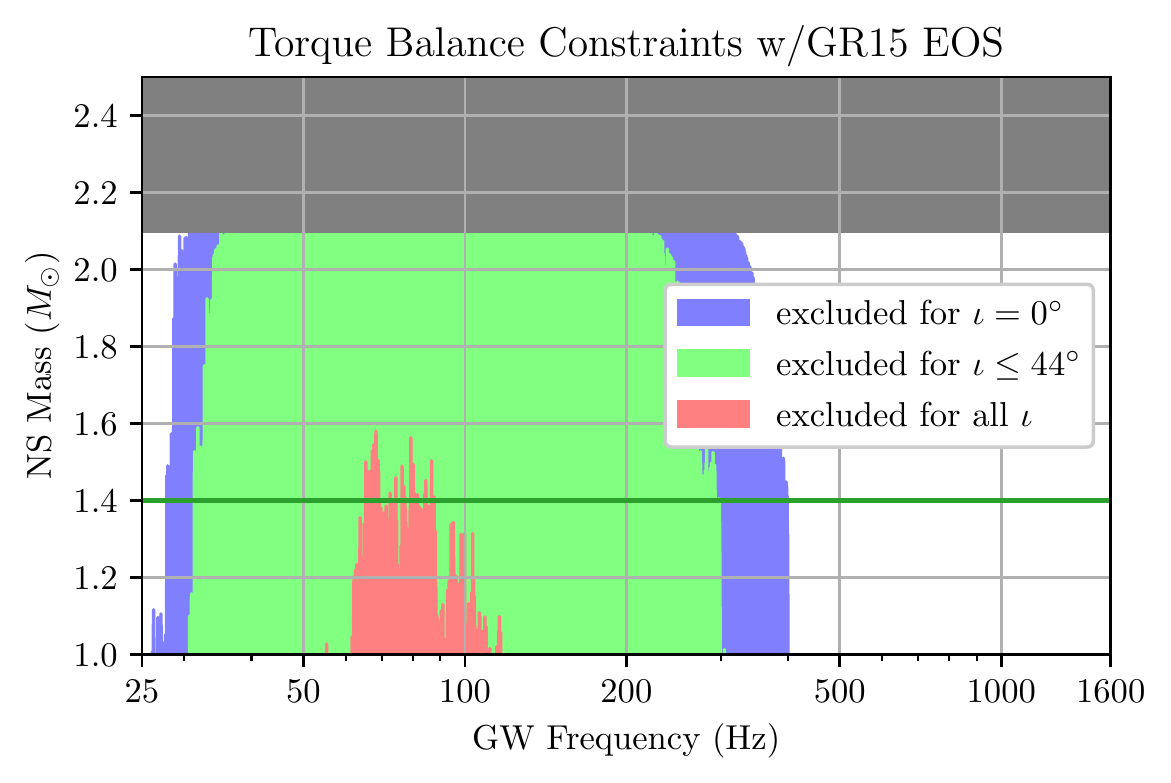}
        }%
        \subfigure[]{%
           \label{fig:viterbi_o3a}
\includegraphics[width=0.4\textwidth]{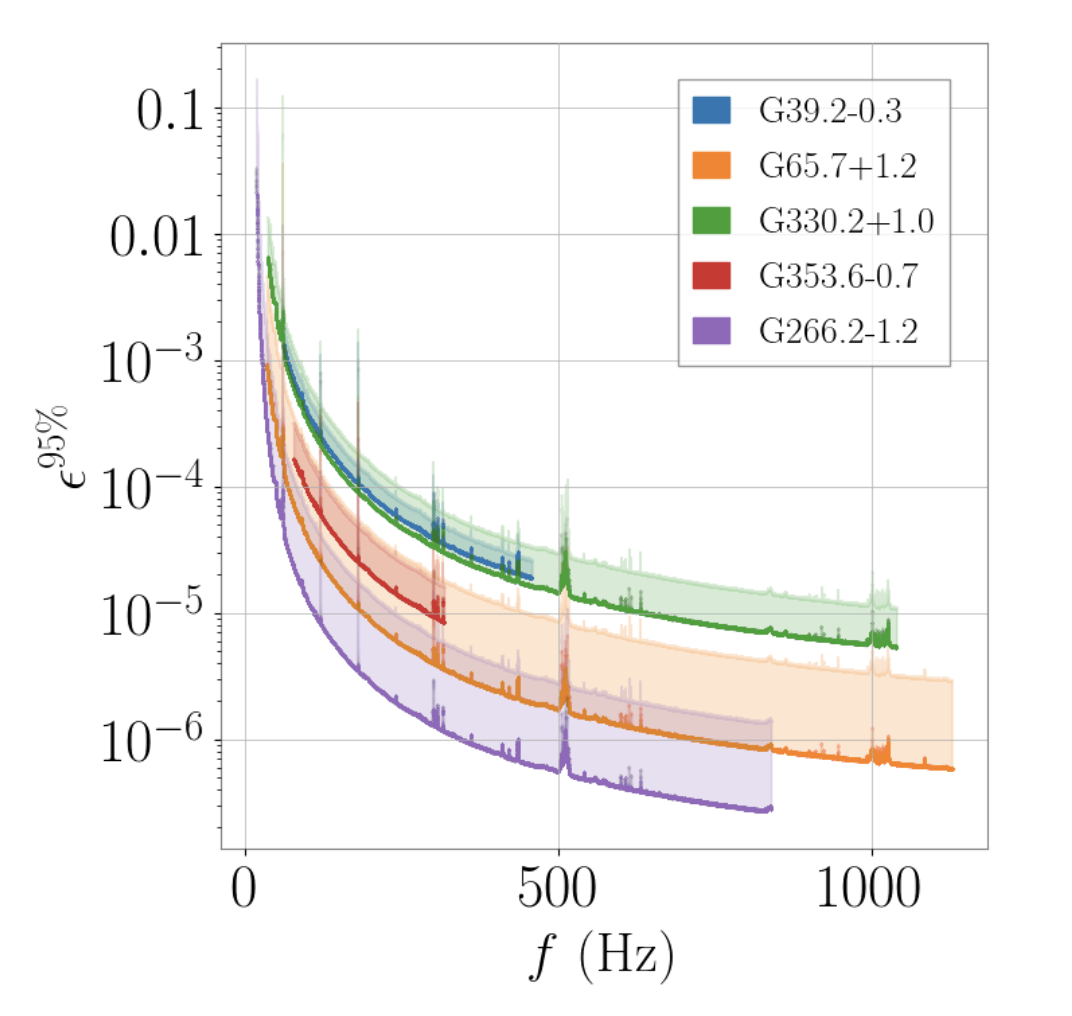}
        }\\ 
    \end{center}
    \caption[]{%
   (a) Exclusion regions of the \NS mass and \GW frequency of Sco X-1, for a given equation of state and inclination angles \cite{LIGOScientific:2022enz}. (b) Ellipticity upper limits for a variety of supernova remnants \cite{LIGOScientific:2021mwx}. }%
     \label{fig:directed}
\end{figure*}

In directed searches for X-ray binaries, such as Sco X-1, we expect that the more luminous the source is, the stronger the \GWs will be. These systems could exist in a state of rotational equilibrium: the spin-down due to \GW is balanced by the spin-up induced by accretion \cite{Bildsten:1998ey}. X-ray binaries are therefore promising sources; however, additional complications enter, namely the uncertainty on the orbital parameters. While $\dot{f}$ is typically not searched over, the orbital Doppler modulation induced in this case severely limits $\TFFT$, because of the sheer number of templates needed to cover the orbital parameter / frequency parameter space. Furthermore, the uncertainty on the orbital parameters becomes larger over the course of the observation run. Searches for Sco X-1 and other low-mass X-ray binaries therefore are very challenging; however, despite these drawbacks, stringent constraints have been placed and, for the first time, and have beaten the so-called torque-balance limit. In Fig. \ref{fig:gr15sco-x1}, constraints on the NS star mass, as a function of \GW freqeuncy, are placed assuming that the source has a different inclination angles with respect to us. Furthermore, we also show, in Fig. \ref{fig:viterbi_o3a}, constraints on the ellipticity as a function of the \GW frequency for some known supernova remnants.

\subsection{All-sky searches}
\label{sec:allsky}
When sky location and source parameters are unknown, we perform \emph{all-sky} searches for isolated neutron stars, or a neutron star in a binary system. Again, semi-coherent techniques must be employed, since matched filtering is too computationally expensive. Furthermore, all-sky searches take an ``eyes wide-open'' approach, and do not rely on any electromagnetic information.

\begin{figure*}[ht!]
     \begin{center}
        \subfigure[ ]{%
            \label{fig:h0_ul}
            \includegraphics[width=0.27\textwidth]{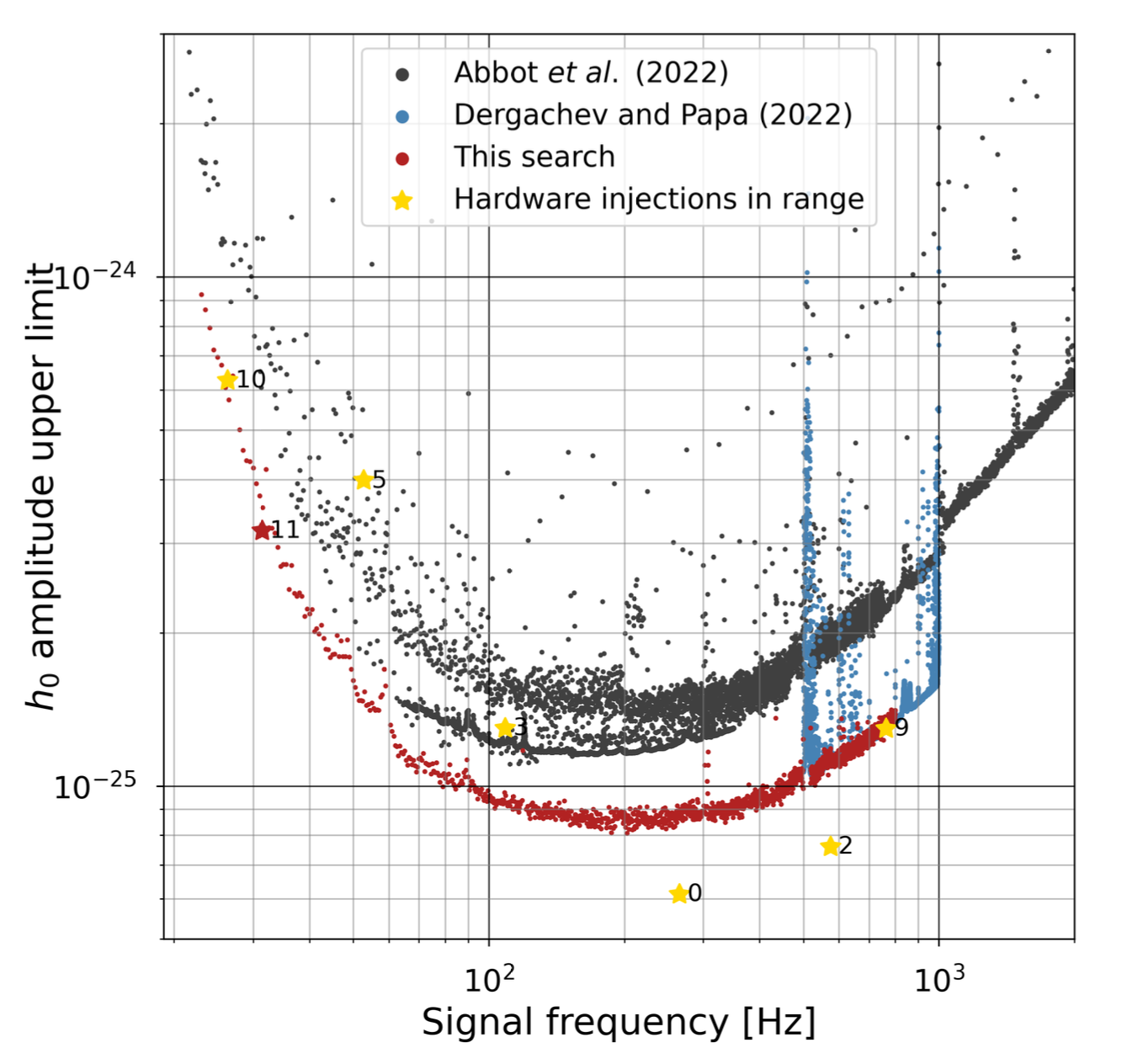}
        }%
                \subfigure[]{%
           \label{fig:gev-excess}
\includegraphics[width=0.33\textwidth]{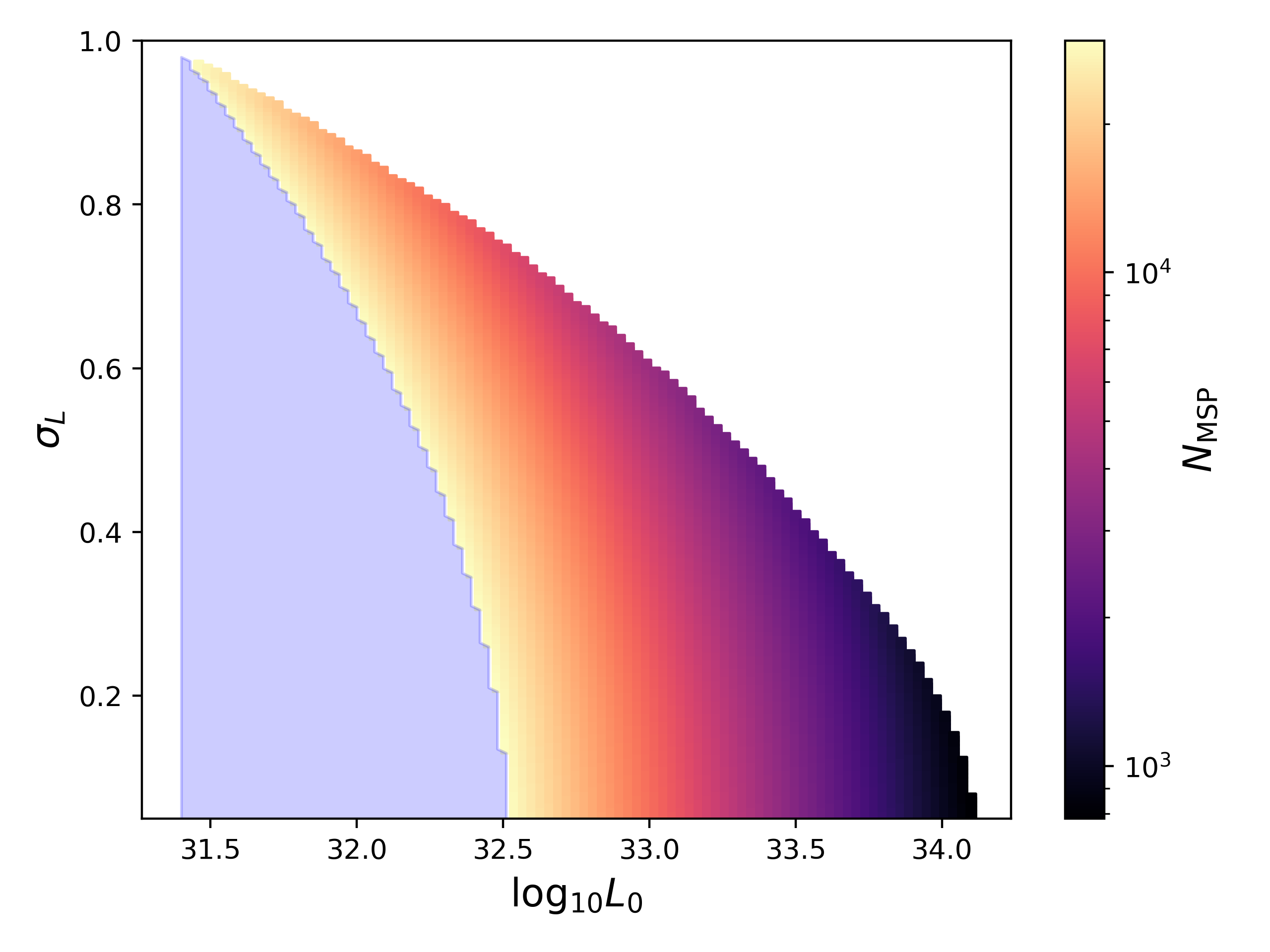}
        }
        \subfigure[]{%
           \label{fig:binaryallsky}
\includegraphics[width=0.33\textwidth]{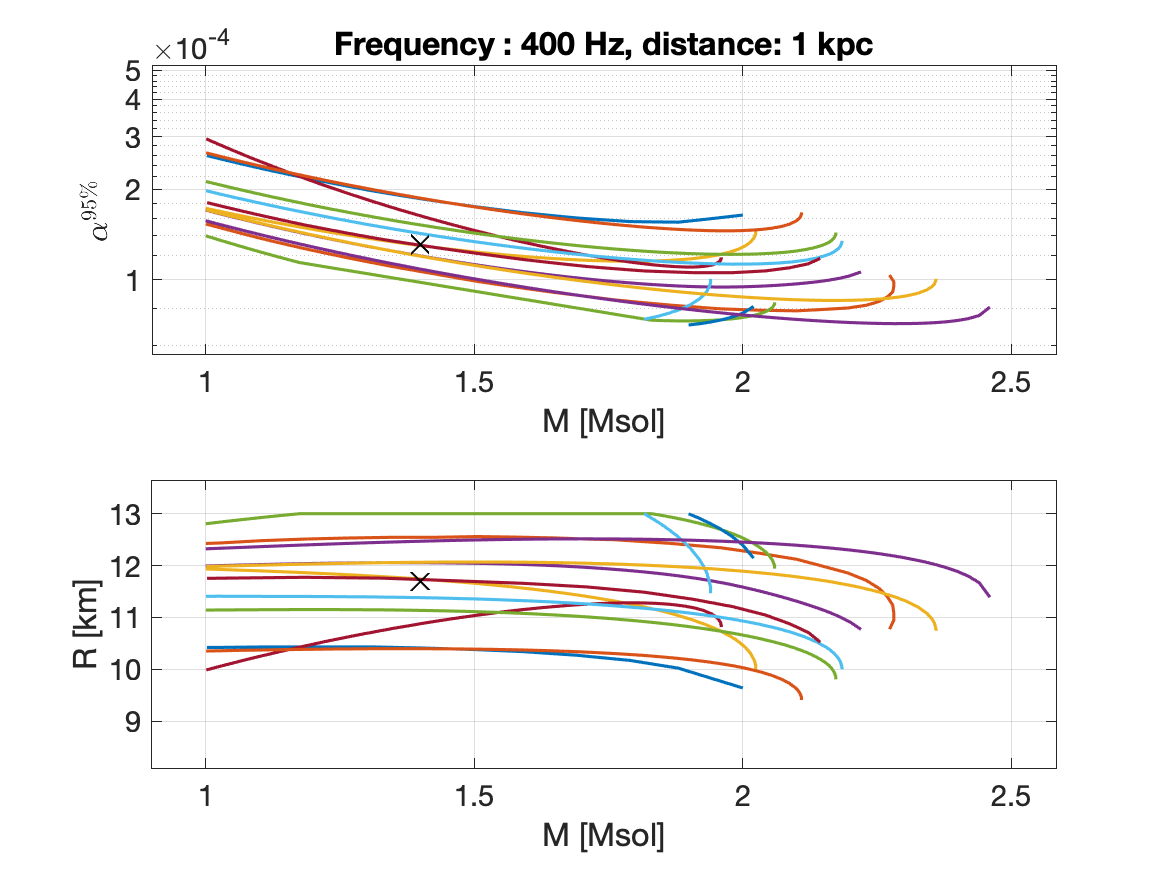}
        }\\ 
    \end{center}
    \caption[]{%
   (a) Upper limits on the strain amplitude at each \GW frequency, averaged over sky location, for all-sky searches inside and outside the LVK \cite{Steltner:2023cfk}. (b) Exclusion regions on the luminosity function parameter space using the results in (a) \cite{Miller:2023qph} (c) Constraints on the r-mode saturation amplitude from an all-sky search for neutron stars in binary systems, for different equations of state \cite{Covas:2022rfg}. }%
     \label{fig:all-sky-gev}
\end{figure*}

In O3, four different LVK methods, based on the Hough Transform, neural networks, Viterbi, and the $\mathcal{F}$-statistic \cite{KAGRA:2022dwb} covered a wide range of the $f_0$/$\dot{f}$ parameter space ($f_0$: [10,2000]; $\dot{f}=[-2,+2]\times 10^{-8}$ Hz/s) with arc-second resolution in the sky, while a similar search was performed with the Einstein\@Home pipeline by external scientists with significantly longer $\TFFT$ and thus even finer sky resolution \cite{Steltner:2023cfk}. All searches resulted in stringent constraints on neutron star ellipticity at different sky locations and frequencies, which are shown in Fig. \ref{fig:h0_ul} .

From the all-sky search results, we can even make a statement about the GeV excess in the galactic center, if it is explained by millisecond pulsars. We can calculate the probability of detecting a neutron star with our search techniques, and then employ a luminosity function and a distribution over the \NS ellipticity to obtain a number of possibly detectable \NS. If that number exceeds one, we can exclude the portion the luminosity function parameters that gave rise to that  predicted number of MSPs to explain the GeV excess, shown in Fig. \ref{fig:gev-excess} \cite{Miller:2023qph}. 

Furthermore, all-sky searches for \NS in binary systems were performed in O3a and O3, which again resulted in constraints on ellipticity \cite{LIGOScientific:2020qhb,Covas:2022rfg}. The searches considered particular frequency bands and certain ranges of orbital parameters that contained most of the parameters of known neutron stars in binary systems. With null results, constraints were derived on the ellipticity and r-mode amplitude for different equation of states, see Fig. \ref{fig:binaryallsky}.

\section{Dark Matter}

\subsection{Gravitational waves from boson clouds around rotating black holes} 

If ultralight dark matter exists and is bosonic, it could accumulate around rotating black holes, and build up a macroscropic ``boson cloud'' over a very short period of time \cite{Brito:2015oca}. This phenomenon, known as ``superradiance'', extracts energy and mass from a black hole. The physics of this cloud can be analogously to a Hydrogen atom, with the exception that there is no limit on the occupation number of bosons in any particular energy state. The lack of this limitation allows the cloud to exponentially increase in size and extract more and more energy from the black hole. Once this process saturates, bosons annihilate into gravitons, the cloud shrinks, and \GWs are emitted from a fixed energy level, meaning that they are monochromatic, with a small frequency drift $\dot{f}$ due to the classical contraction of the cloud over time.

\begin{figure*}[ht!]
     \begin{center}
        \subfigure[ ]{%
            \label{fig:boson_cloud}
            \includegraphics[width=0.3\textwidth]{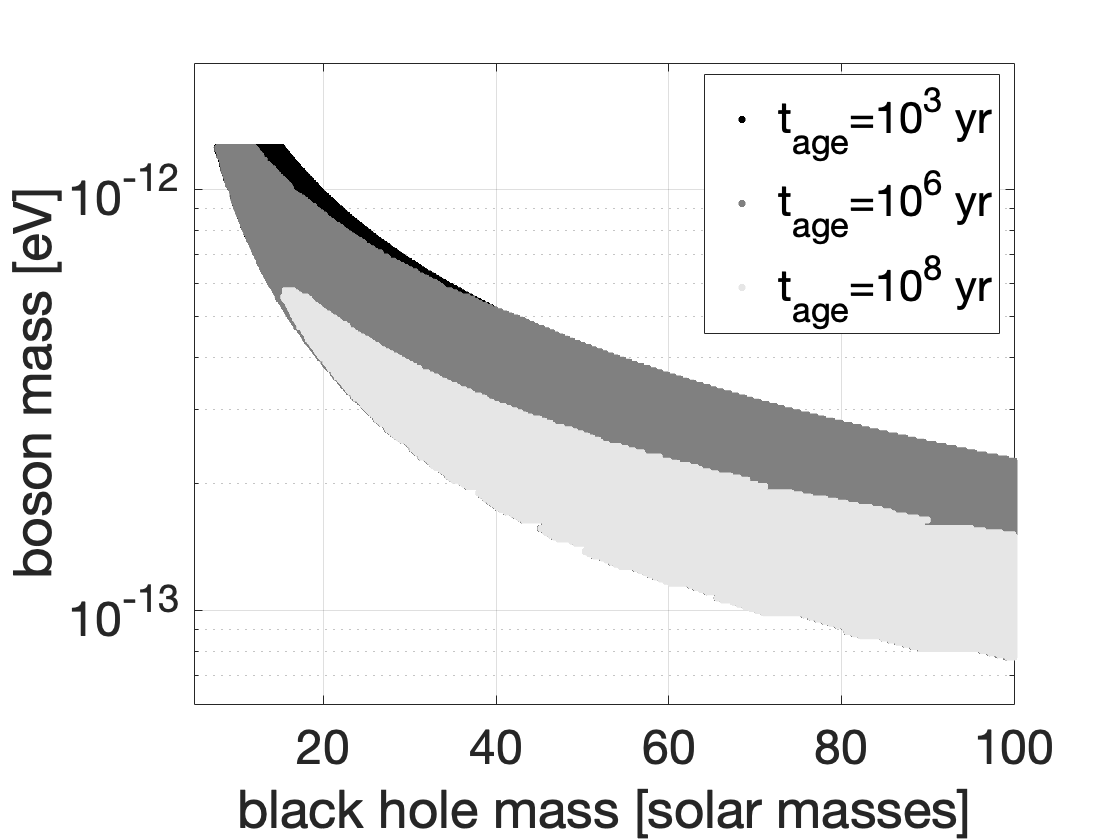}
        }%
        \subfigure[]{%
           \label{fig:direct_dm}
\includegraphics[width=0.3\textwidth]{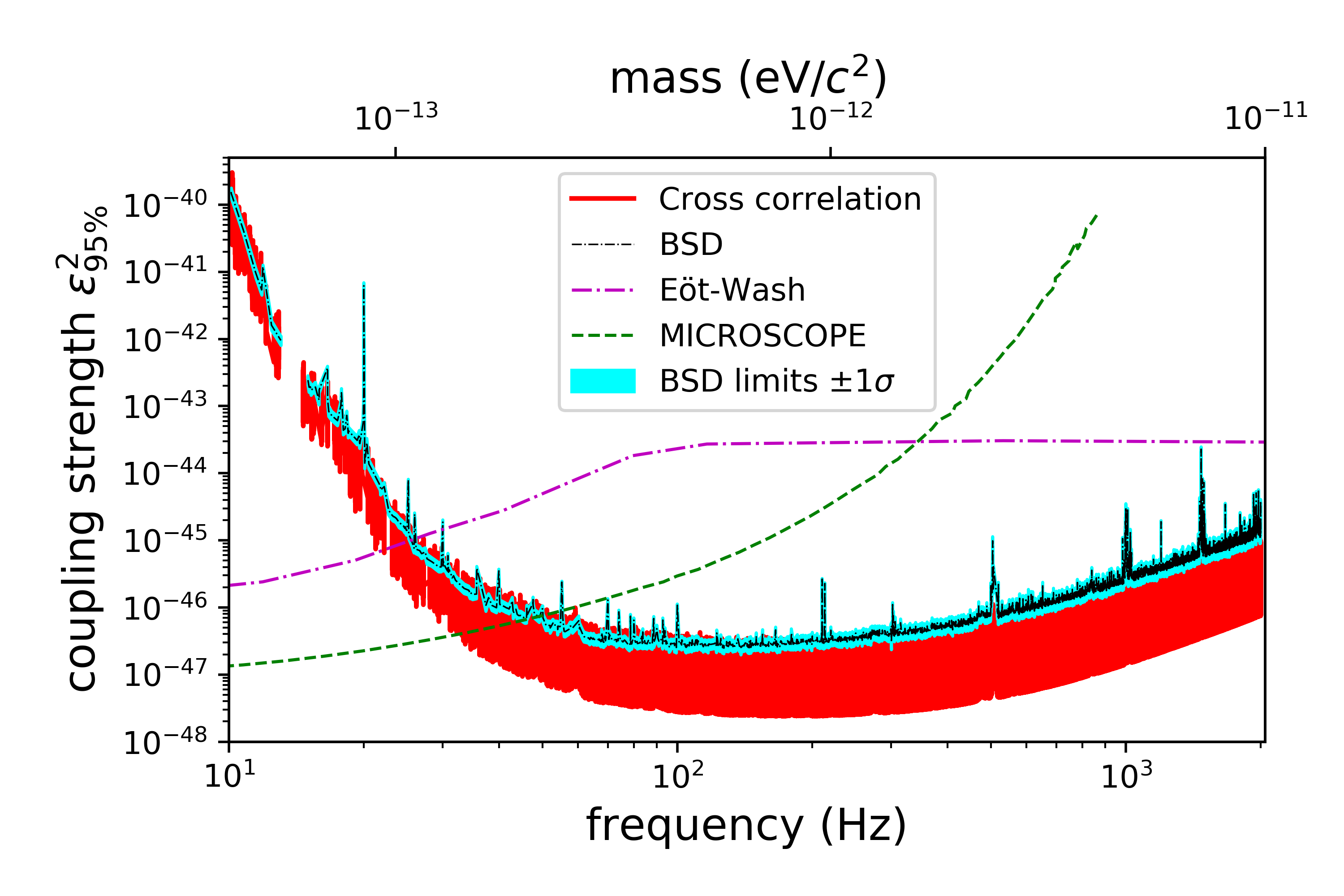}
        }
                \subfigure[]{%
           \label{fig:pbh-lims}
\includegraphics[width=0.3\textwidth]{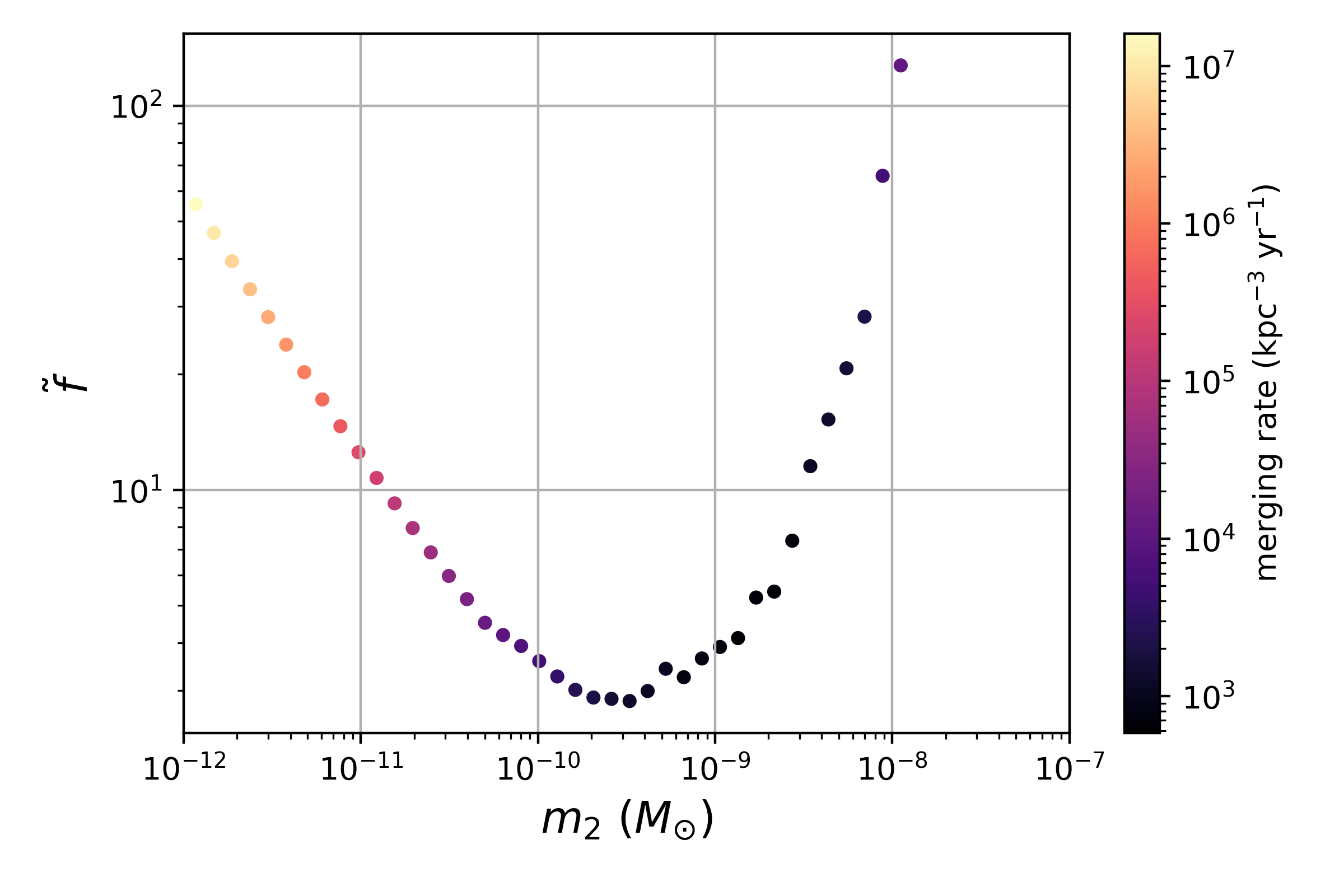}
        }\\ 
    \end{center}
    \caption[]{%
   (a) Exclusion regions for the boson/black hole mass parameter space for different aged systems, averaged over sky location, 1 kpc away \cite{LIGOScientific:2021rnv}. (b) Upper limits on the coupling strength of dark photons to baryons in the LIGO mirrors \cite{LIGOScientific:2021odm}. (c) Upper limit on the fraction of \DM that \PBHs could compose as a function of $m_2$ in an asymmetric mass-ratio binary system, in each $m_1=2.5M_\odot$, calculating using the limits in Fig. \ref{fig:h0_ul} \cite{KAGRA:2022dwb}. }%
     \label{fig:dm}
\end{figure*}

In O3, an all-sky search was performed for scalar boson clouds around rotating black holes, resulting in exclusions of a combination of boson mass/black hole mass from existing at a fixed distance away from us and with a fixed spin \cite{LIGOScientific:2021rnv}, as shown in Fig. \ref{fig:boson_cloud}. This all-sky search differs from those discussed in Sec. \ref{sec:allsky} in that the $\dot{f}$ parameter is not searched over, implying a reduced penalty due to the trials factor, and that there is explicit consideration of deviations from the ``exactly monochromatic'' signal model, via a moving average in the time/frequency map that considers that a particular $\TFFT$ may not capture all the unknown signal modulations.

\subsection{Direct Detection}

Ultralight \DM particles could couple to the standard model particles in \GW interferometers, and imprint a detectable signal \cite{Pierce:2018xmy}. While this effect is not due to a passing \GW, the kind of signal we expect is monochromatic and persistent, with stochastic frequency fluctuations induced by the earth's motion relative to the incoming \DM field. In essence, the \GW interferometers sit in a sea of \DM that oscillates the components of the interferometer at an almost fixed frequency. The stochastic frequency fluctuations are of $\mathcal{O}(10^{-6})$, a couple of orders of magnitude smaller than that induced on a \NS signal from the Doppler motion of the earth relative to the source \cite{Miller:2020vsl}.

In O3, a search for dark photon dark matter, a type of particle that would couple to baryons or baryon-lepton number in each LIGO/Virgo mirror, was performed, resulting in the most stringent constraints for masses of $\sim 5\times 10^{-13}-10^{-11}$ eV/$c^2$, and that surpassed upper limits from existing \DM experiments (MICROSCOPE and Eot-Wash) by a few orders of magnitude \cite{LIGOScientific:2021odm}. These limits are shown in Fig. \ref{fig:direct_dm}. 

\subsection{Primordial black hole binaries}

Because we assume in all-sky searches a simple signal model, the same all-sky search results in Fig. \ref{fig:h0_ul} can be interpreted as upper limits on the fraction of dark matter that asteroid-mass primordial black holes could compose, $f_{\rm pbh}$ if they were inspiraling towards each other. In essence, at such low masses, the frequency evolution of the \GW signal does not ``chirp'' like the detected BBH and BNS mergers, but is in fact quasi-monochromatic, and the system will not merge for billions of years \cite{Miller:2021knj}. Fig. \ref{fig:pbh-lims} shows a model-independent parameter $\tilde{f}$ that relates to $f_{\rm pbh}$ as a function of the secondary mass of an inspiralling binary system, in which the primary mass is of $m_1=2.5M_{\odot}$. $\tilde{f}$ is greater than 1, meaning that we cannot yet place physical upper limits on $f_{\rm pbh}$; however, as the sensitivity improves in the next observing runs, we should have the first stringent constraints in asteroid-mass regime \cite{KAGRA:2022dwb}.

\section{Conclusions and Prospects}

Searches for \CWs are able to probe a wide range of multi-messenger astrophysics employing a very simple signal model. Ultralight dark matter, primordial black holes, isolated neutron stars, neutron stars in X-ray binaries -- all of these sources can be probed with the same ways of thinking presented in these proceedings. The breadth of \CW sources is only matched by the variety of methods that have been designed to search for them: computational expense, the non-Gaussian noise, and the prospects of uncertain \NS and \DM physics have driven scientists to think of new ways to tackle the ``simple'' problem of finding sinusoids in time-series data. 

The O3 run brought unprecedented sensitivity towards known millisecond pulsars, dark matter and Scorpius X-1, and served as the first observing run in which LVK-wide searches for \DM were performed. \CW searches are expanding their horizons beyond looking for``just'' known and unknown neutron stars, but are beginning to build even more multi-messenger bridges between \GW astronomy and cosmology, particle physics and tests of fundamental physics.

The next observing run will likely yield improved sensitivity towards each of these sources. The future for \CWs is bright, and will hopefully feature a detection. But even without one, the breadth of \CW physics, coupled with the applicability of \CW methods to a variety of problems outside of \CWs, makes investing and supporting \CW science worthwhile.

\section*{References}\small

\bibliography{biblio}





\end{document}